\documentclass{aastex6}

\fullcollaborationName{The Friends of AASTeX Collaboration}

\begin{document}



\title{CALET Upper Limits on X-ray and Gamma-ray Counterparts of GW 151226\footnote{Corresponding authors: T. Sakamoto, S. Nakahira and Y. Asaoka}}


\author{O. Adriani\altaffilmark{1,40,41}, 
Y. Akaike\altaffilmark{2,3}, 
K. Asano\altaffilmark{4}, 
Y. Asaoka\altaffilmark{5,6}, 
M.G. Bagliesi\altaffilmark{7,40}, 
G. Bigongiari\altaffilmark{7,40}, 
W.R. Binns\altaffilmark{8}, \\
S. Bonechi\altaffilmark{7,40},  
M. Bongi\altaffilmark{1,40,41},
P. Brogi\altaffilmark{7,40}, 
J.H. Buckley\altaffilmark{8}, 
N. Cannady\altaffilmark{9},
G. Castellini\altaffilmark{1,40,41},  
C. Checchia\altaffilmark{10}, \\
M.L. Cherry\altaffilmark{9}, 
G. Collazuol\altaffilmark{10}, 
V. Di Felice\altaffilmark{11,40}, 
K. Ebisawa\altaffilmark{12}, 
H. Fuke\altaffilmark{12}, 
T.G. Guzik\altaffilmark{9}, 
T. Hams\altaffilmark{13,3}, \\
M. Hareyama\altaffilmark{14},  
N. Hasebe\altaffilmark{6}, 
K. Hibino\altaffilmark{15}, 
M. Ichimura\altaffilmark{16}, 
K. Ioka\altaffilmark{17}, 
W. Ishizaki\altaffilmark{4}, 
M.H. Israel\altaffilmark{8}, 
A. Javaid\altaffilmark{9}, \\
K. Kasahara\altaffilmark{6}, 
J. Kataoka\altaffilmark{6}, 
R. Kataoka\altaffilmark{18}, 
Y. Katayose\altaffilmark{19}, 
C. Kato\altaffilmark{20}, 
N. Kawanaka\altaffilmark{21}, 
Y. Kawakubo\altaffilmark{22}, \\
H. Kitamura\altaffilmark{23}, 
H.S. Krawczynski\altaffilmark{8}, 
J.F. Krizmanic\altaffilmark{2,3}, 
S. Kuramata\altaffilmark{16}, 
T. Lomtadze\altaffilmark{24}, 
P. Maestro\altaffilmark{7,40}, \\
P.S. Marrocchesi\altaffilmark{7,40}, 
A.M. Messineo\altaffilmark{24}, 
J.W. Mitchell\altaffilmark{25}, 
S. Miyake\altaffilmark{26}, 
K. Mizutani\altaffilmark{27}, 
A.A. Moiseev\altaffilmark{38,3}, 
K. Mori\altaffilmark{6,12},  \\
M. Mori\altaffilmark{28}, 
N. Mori\altaffilmark{1,40,41},  
H.M. Motz\altaffilmark{39}, 
K. Munakata\altaffilmark{20}, 
H. Murakami\altaffilmark{6}, 
Y.E. Nakagawa\altaffilmark{12}, 
S. Nakahira\altaffilmark{5}, \\
J. Nishimura\altaffilmark{12}, 
S. Okuno\altaffilmark{15}, 
J.F. Ormes\altaffilmark{29}, 
S. Ozawa\altaffilmark{6}, 
L. Pacini\altaffilmark{1,40,41},
F. Palma\altaffilmark{11,40}, 
P. Papini\altaffilmark{1,40,41}, \\
A.V. Penacchioni\altaffilmark{7, 37}, 
B.F. Rauch\altaffilmark{8}, 
S. Ricciarini\altaffilmark{1,40,41}, 
K. Sakai\altaffilmark{13,3}, 
T. Sakamoto\altaffilmark{22}, 
M. Sasaki\altaffilmark{38,3}, 
Y. Shimizu\altaffilmark{15}, \\
A. Shiomi\altaffilmark{30}, 
R. Sparvoli\altaffilmark{11,40},  
P. Spillantini\altaffilmark{1,40,41},  
F. Stolzi\altaffilmark{7,40},
I. Takahashi\altaffilmark{31}, 
M. Takayanagi\altaffilmark{12}, 
M. Takita\altaffilmark{4}, \\
T. Tamura\altaffilmark{15}, 
N. Tateyama\altaffilmark{15}, 
T. Terasawa\altaffilmark{32}, 
H. Tomida\altaffilmark{12}, 
S. Torii\altaffilmark{5,6}, 
Y. Tsunesada\altaffilmark{33}, 
Y. Uchihori\altaffilmark{23}, 
S. Ueno\altaffilmark{12}, \\
E. Vannuccini\altaffilmark{1,40,41}, 
J.P. Wefel\altaffilmark{9}, 
K. Yamaoka\altaffilmark{34}, 
S. Yanagita\altaffilmark{35}, 
A. Yoshida\altaffilmark{22}, 
K. Yoshida\altaffilmark{36}, and 
T. Yuda\altaffilmark{4}} 
\affil{$^{1}$University of Florence, Via Sansone, 1 - 50019 Sesto, Fiorentino, Italy\\
$^{2}$Universities Space Research Association, 7178 Columbia Gateway Drive Columbia, MD 21046, USA.\\
$^{3}$CRESST and Astroparticle Physics Laboratory NASA/GSFC, Greenbelt, MD 20771, USA\\
$^{4}$Institute for Cosmic Ray Research, The University of Tokyo, 5-1-5 Kashiwa-no-Ha, Kashiwa, Chiba 277-8582, Japan\\
$^{5}$JEM Mission Operations and Integration Center, Human Spaceflight Technology Directorate, 
Japan Aerospace Exploration Agency, 2-1-1 Sengen, Tsukuba, Ibaraki 305-8505, Japan\\
$^{6}$Research Institute for Science and Engineering, Waseda University, 3-4-1 Okubo, Shinjuku, Tokyo 169-8555, Japan\\
$^{7}$University of Siena, Rettorato, via Banchi di Sotto 55, 53100 Siena, Italy\\
$^{8}$Department of Physics,Washington University, One Brookings Drive, St. Louis, MO 63130-4899, USA\\
$^{9}$Department of Physics and Astronomy, Louisiana State University, 202 Nicholson Hall, Baton Rouge, LA 70803, USA\\
$^{10}$Department of Physics and Astronomy, University of Padova, Via Marzolo, 8, 35131 Padova, Italy\\
$^{11}$University of Rome Tor Vergata, Via della Ricerca Scientifica 1, 00133 Rome, Italy\\
$^{12}$Institute of Space and Astronautical Science, Japan Aerospace Exploration Agency, 
3-1-1 Yoshinodai, Chuo, Sagamihara, Kanagawa 252-5210, Japan\\
$^{13}$Department of Physics, University of Maryland, Baltimore County, 1000 Hilltop Circle, Baltimore, MD 21250, USA\\
$^{14}$St. Marianna University School of Medicine, 2-16-1, Sugao, Miyamae-ku, Kawasaki, Kanagawa 216-8511, Japan\\
$^{15}$Kanagawa University, 3-27-1 Rokkakubashi, Kanagawa, Yokohama, Kanagawa 221-8686, Japan\\
$^{16}$Faculty of Science and Technology, Graduate School of Science and Technology, Hirosaki University, 
3, Bunkyo, Hirosaki, Aomori 036-8561, Japan\\
${17}$Yukawa Institute for Theoretical Physics, Kyoto University, Kitashirakawa Oiwakecho, Sakyo, Kyoto 606-8502, Japan\\
$^{18}$National Institute of Polar Research, 10-3, Midori-cho, Tachikawa, Tokyo 190-8518, Japan\\
$^{19}$Faculty of Engineering, Division of Intelligent Systems Engineering, Yokohama National University, 
79-5 Tokiwadai, Hodogaya, Yokohama 240-8501, Japan\\
$^{20}$Faculty of Science, Shinshu University, 3-1-1 Asahi, Matsumoto, Nagano 390-8621, Japan\\
$^{21}$School of Science, The University of Tokyo, 7-3-1 Hongo, Bunkyo, Tokyo 113-003, Japan\\
$^{22}$College of Science and Engineering, Department of Physics and Mathematics, Aoyama Gakuin University,  
5-10-1 Fuchinobe, Chuo, Sagamihara, Kanagawa 252-5258, Japan\\
$^{23}$National Institute of Radiological Sciences, 4-9-1 Anagawa, Inage, Chiba 263-8555, Japan\\
$^{24}$University of Pisa and INFN, Italy\\
$^{25}$Astroparticle Physics Laboratory, NASA/GSFC, Greenbelt, MD 20771, USA\\
$^{26}$Department of Electrical and Electronic Systems Engineering, National Institute of Technology, Ibaraki College, 866 Nakane, 
Hitachinaka, Ibaraki 312-8508 Japan\\
$^{27}$Saitama University, Shimo-Okubo 255, Sakura, Saitama, 338-8570, Japan\\
$^{28}$Department of Physical Sciences, College of Science and Engineering, Ritsumeikan University, Shiga 525-8577, Japan\\
$^{29}$Department of Physics and Astronomy, University of Denver, Physics Building, Room 211, 2112 East Wesley Ave., Denver, CO 
80208-6900, USA\\
$^{30}$College of Industrial Technology, Nihon University, 1-2-1 Izumi, Narashino, Chiba 275-8575, Japan\\
$^{31}$Kavli Institute for the Physics and Mathematics of the Universe, The University of Tokyo, 5-1-5 Kashiwanoha, Kashiwa, 277-8583, Japan\\
$^{32}$RIKEN, 2-1 Hirosawa, Wako, Saitama 351-0198, Japan\\
$^{33}$Division of Mathematics and Physics, Graduate School of Science, Osaka City University, 3-3-138 Sugimoto, Sumiyoshi, 
Osaka 558-8585, Japan\\
$^{34}$Nagoya University, Furo, Chikusa, Nagoya 464-8601, Japan\\
$^{35}$Graduate School of Science and Engineering, Ibaraki University, 2-1-1 Bunkyo, Mito, Ibaraki 310-8512, Japan\\
$^{36}$Department of Electronic Information Systems, Shibaura Institute of Technology, 307 Fukasaku, Minuma, Saitama 337-8570, Japan\\
$^{37}$ASI Science Data Center (ASDC), Via del Politecnico snc, 00133 Rome, Italy\\
$^{38}$Department of Astronomy, University of Maryland, College Park, Maryland 20742, USA \\ 
$^{39}$International Center for Science and Engineering Programs, Waseda University, 3-4-1 Okubo, Shinjuku, Tokyo 169-8555, Japan\\
$^{40}$National Institute for Nuclear Physics (INFN), Piazza dei Caprettari, 70 - 00186 Rome, Italy\\
$^{41}$Institute of Applied Physics (IFAC),  National Research Council (CNR), Via Madonna del Piano, 10, 50019 Sesto, Fiorentino, Italy\\
}

\begin{abstract}
We present upper limits in the hard X-ray and gamma-ray bands at the time of the LIGO gravitational-wave event 
GW 151226 derived from the CALorimetric Electron Telescope ({\it CALET}) observation.  The main instrument of 
CALET, CALorimeter (CAL), observes gamma-rays from $\sim$1 GeV up to 10 TeV with a field of view of $\sim$2 sr.  
The {\it CALET} gamma-ray burst monitor (CGBM) views $\sim$3 sr and $\sim$2$\pi$ sr of the 
sky in the 7 keV - 1 MeV and the 40 keV - 20 MeV bands, respectively, by using two different scintillator-based instruments.  
The CGBM covered 32.5\% and 49.1\% of the GW 151226 sky localization probability in the 7 keV - 1 MeV and 40 keV - 20 MeV 
bands respectively.  
We place a 90\% upper limit of $2 \times 10^{-7}$ erg cm$^{-2}$ s$^{-1}$ in the 1 - 100 GeV band where CAL reaches 
15\% of the integrated LIGO probability ($\sim$1.1 sr).  
The CGBM 7 $\sigma$ upper limits are $1.0 \times 10^{-6}$ erg cm$^{-2}$ s$^{-1}$ 
(7-500 keV) and $1.8 \times 10^{-6}$ erg cm$^{-2}$ s$^{-1}$ (50-1000 keV) for one second exposure.   
Those upper limits correspond to the luminosity of $3$-$5$ $\times 10^{49}$ erg s$^{-1}$
which is significantly lower than typical short GRBs.  
\end{abstract}




\section{Introduction}\label{sec:intro}



The first gravitational-wave detection by the Laser Interferometer Gravitational-Wave 
Observatory (LIGO) on GW 150914 confirmed the existence not only of gravitational 
waves from astronomical objects but also of a binary black hole system with several tens of solar 
masses \citep{ligo_gw150914}.  Based solely on the gravitational-wave signals  
recorded by two LIGO detectors, the current hypothesis is that GW 150914 was the result of a merger 
of two black holes with initial masses of $36_{-4}^{+5} M_{\sun}$ and $29_{-4}^{+4} M_{\sun}$ 
at the luminosity distance of $410_{-180}^{+160}$ Mpc.  The {\it Fermi} Gamma-ray 
Burst Monitor (Fermi-GBM) reported a possible weak gamma-ray transient source above 50 keV 
at 0.4 s after the GW 150914 trigger \citep{fermigbm_gw150914}.  However, 
the upper limit provided by the {\it INTEGRAL} ACS instrument 
in a gamma-ray energy band similar to the {\it Fermi}-GBM energy band 
is not consistent with a possible gamma-ray 
counterpart of GW 150914 suggested by the {\it Fermi}-GBM \citep{integral_gw150914}.  
No electromagnetic counterpart of GW 150914 was found in radio, optical, near-infrared, 
X-ray and high-energy gamma-ray \citep{em_gw150914}.  

GW 151226 (LIGO-Virgo trigger ID: G211117) is the 2nd gravitational-wave candidate 
identified by both LIGO Hanford Observatory and LIGO Livingston Observatory with a 
high significance (the false alarm rate of less than one per 1000 years by the on-line 
search) at 3:38:53.647 UT 
on December 26, 2015 \citep{ligo_gw151226}.  According to a Bayesian 
parameter estimation analysis, the event is very likely a binary black hole merger 
with initial black hole masses of $14.2_{-3.7}^{+8.3} M_{\sun}$ and $7.5_{-2.3}^{+2.3} M_{\sun}$ , 
and final black hole mass of $20.8_{-1.7}^{+6.1} M_{\sun}$ \citep{ligo_O1run}.  The luminosity distance 
of the source is estimated as $440_{-190}^{+180}$ Mpc which corresponds to a redshift of $0.09_{-0.04}^{+0.03}$.   
As far as the electromagnetic counterpart search of GW 151226 in the gamma-ray regime is concerned, 
{\it Fermi}-GBM \citep{fgbm_gw151226}, {\it Fermi} Large Area Telescope (LAT) \citep{flat_gw151226}, 
High-Altitude Water Cherenkov Observatory (HAWC) \citep{hawc_gw151226}, and {\it Astrosat}-CZTI \citep{astrosat_czti_gw151226} 
reported no detections around the GW trigger time.  
According to \citet{racusin2016}, the flux upper limit of {\it Fermi}-GBM is 
from $4.5$ $\times$ 10$^{-7}$ erg cm$^{-2}$ s$^{-1}$ to $9$ $\times$ 10$^{-7}$ erg cm$^{-2}$ s$^{-1}$ 
in the 10-1000 keV band.  
The {\it Fermi}-LAT flux upper limit using the first orbit data after the LIGO trigger is from $2.6 \times 10^{-10}$ erg 
cm$^{-2}$ s$^{-1}$ to $7.8 \times 10^{-9}$ erg cm$^{-2}$ s$^{-1}$ in the 0.1-1 GeV band.  

The CALorimetric Electron Telescope ({\it CALET}; \citet{torii_2015, asaoka_2015}) mission, which was 
successfully launched and emplaced on the Japanese Experiment Module - Exposed Facility (JEM-EF) of 
the International Space Station (ISS) 
in August 2015, was fully operational at the time of GW 151226.  
{\it CALET} consists of two scientific instruments.  The Calorimeter (CAL) is 
the main instrument which is capable of observing high energy electrons 
from $\sim$1 GeV to $\sim$20 TeV, protons, helium and heavy nuclei 
from $\sim$10 GeV to 1000 TeV, and gamma-rays from $\sim$1 GeV 
to $\sim$10 TeV.  The field of view (FOV) of CAL is $\sim$$45^{\circ}$ from the 
zenith direction.  Another instrument, {\it CALET} Gamma-ray Burst Monitor 
(CGBM; \citet{yamaoka2013}), is a gamma-ray burst (GRB) monitor using two different kind 
of scintillators (LaBr$_{3}$(Ce) and BGO) to achieve a broad energy coverage.  
The Hard X-ray Monitor (HXM) using LaBr$_{3}$(Ce) 
covers the energy range from 7 keV up to 1 MeV, and two identical 
modules are equipped in the same direction in {\it CALET}.  
The Soft Gamma-ray Monitor (SGM) based on BGO covers the energy range from 40 keV to 20 MeV.  
The FOV of HXM and SGM are $\sim$$60^{\circ}$ and $\sim$$110^{\circ}$ from the boresight respectively.  
The CGBM has been detecting GRBs at an average rate of 3-4 events per month.  

Around the trigger time of GW 151226, {\it CALET} was performing regular scientific 
data collection.  Between 3:30 and 3:43 UT, the CAL was operating in the low-energy gamma-ray mode, which is 
an operation mode with a lower energy threshold of 1 GeV.  
The high voltages of CGBM were 
set at the nominal values around 3:20 UT and turned off around 3:40 UT to avoid 
high background radiation area.  There was no CGBM on-board trigger at the trigger time of GW 151226.  

\section{Observation}\label{sec:obs}

\subsection{CGBM Data Analysis and Results}\label{subsec:cgbm}

\begin{figure}
\centerline{
\includegraphics[scale=0.4,angle=0]{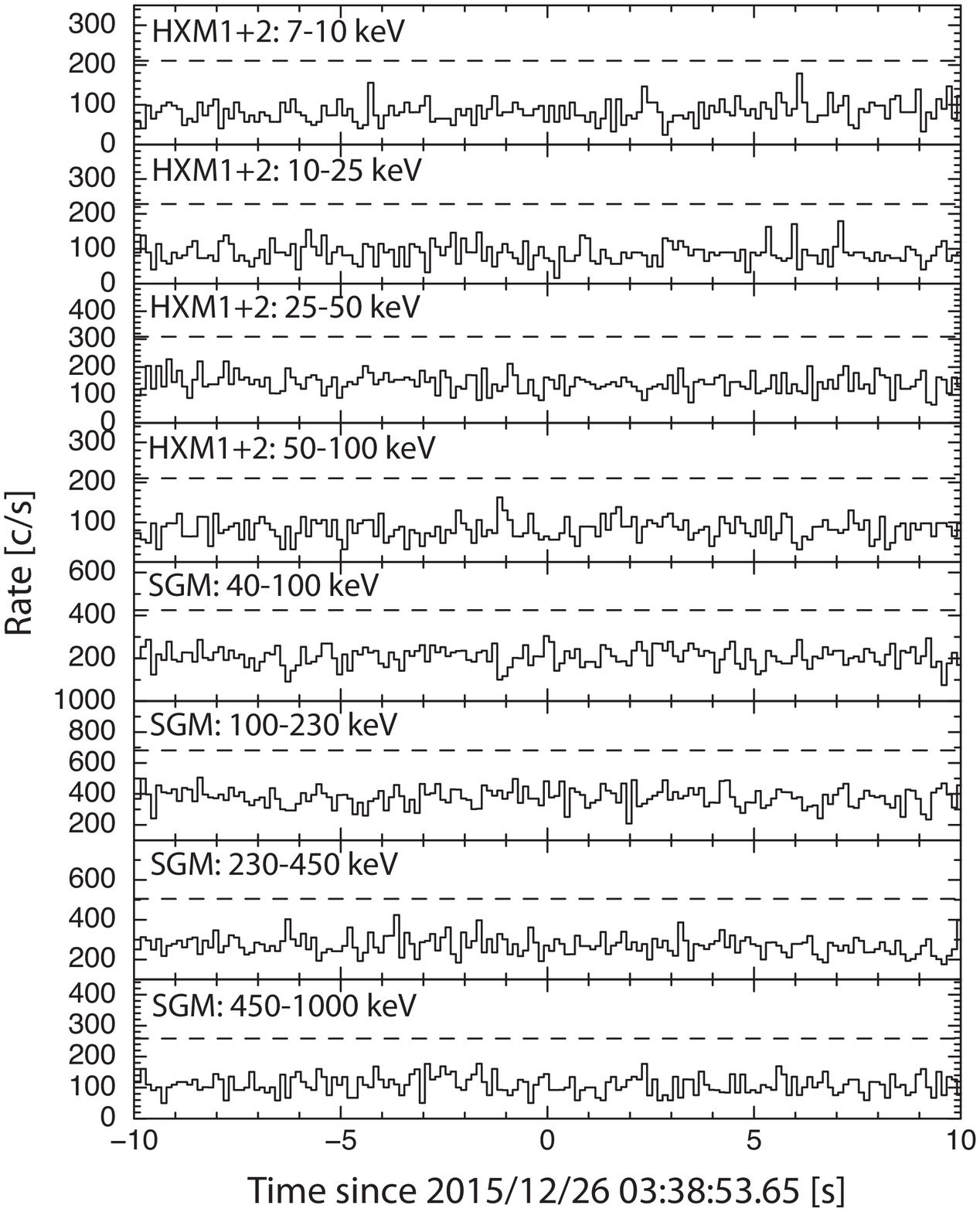}
\includegraphics[scale=0.4,angle=0]{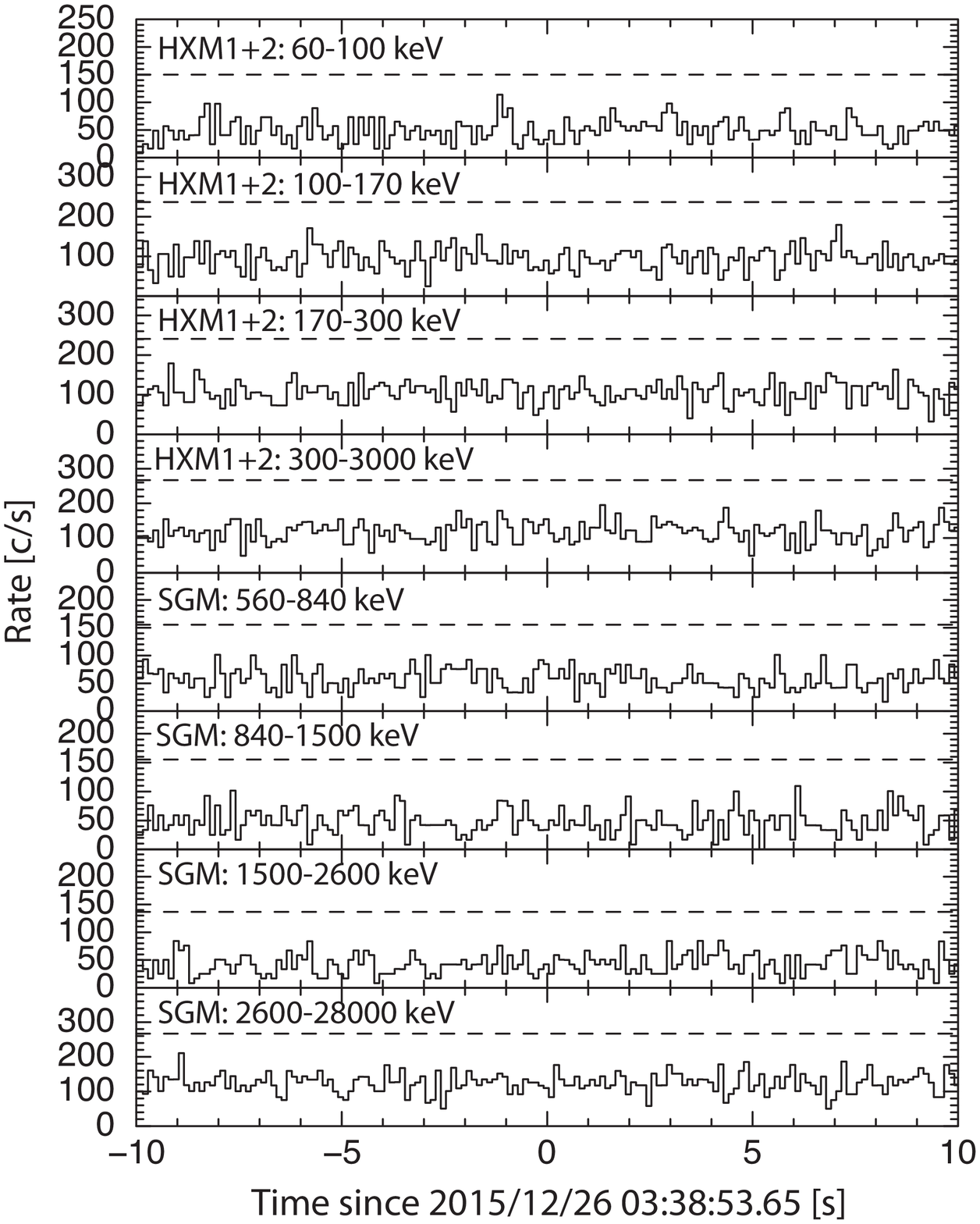}}
\caption{The CGBM light curves in 0.125 s time resolution for the high-gain data (left) and the low-gain 
data (right).  The time is offset from the LIGO trigger time of GW 151226.  The dashed-lines correspond to 
the 5 $\sigma$ level from the mean count rate using the data of $\pm$10 s.}
\label{fig:cgbm_lc}
\end{figure}

At 3:38 UT, the CGBM was operating in nominal operational mode 
in which continuous light curve data in 0.125 s 
time resolution were recorded at eight different energy bands for each instrument.  
The boresight directions of HXM and SGM were (R.A., Dec.) (J2000) 
= (35.6$^{\circ}$, $-28.0^{\circ}$) and (43.5$^{\circ}$, $-22.1^{\circ}$) at the onset of GW 151226.  
Around the GW 151226 event time, no CGBM on-board trigger occurred.  Therefore, the available data to investigate 
the possible counterpart are the continuous light curves mentioned above.  If there is the on-board trigger, 
the time-tagged event data with 62.5 $\mu$s resolution will be generated.  
Figure \ref{fig:cgbm_lc} shows the light curves  
in the 0.125 s time bins in the time range between $\pm$10 s from the GW 151226 trigger time.  As seen in 
the figure, no significant excess is seen in the CGBM data around the trigger time.  We calculate the signal-to-noise 
ratio (SNR) in sliding the time bins of the light curves by selecting the background interval as 8 s, 16 s, 32 s and 64 s 
and the foreground interval as 0.125 s, 0.25 s, 0.5 s, 1 s and 4 s.  
The SNR is calculated as, 
${\rm SNR} = \{N_{\rm fg} - (N_{\rm bg} \Delta t_{\rm fg}/\Delta t_{\rm bg})\}/\sqrt{N_{\rm bg} \Delta t_{\rm fg}/\Delta t_{\rm bg}},$
where $N_{\rm fg}$ is the counts in the foreground interval, $\Delta t_{\rm fg}$ is the integration time of the 
foreground interval, $N_{\rm bg}$ is the counts in the background interval and $\Delta t_{\rm bg}$ is the 
integration time of the background interval.  
The background interval is always prior to the 
foreground interval and there is no time gap between the background and the foreground interval.  
We searched the light curve data for finding signals of individual instruments (HXM1, HXM2 and SGM) 
and the sum of the HXM1 and the HXM2.  
The searched energy bands are all the combinations of 7-10 keV, 10-25 keV, 25-50 keV, 50-100 keV for the high-gain 
data and 60-100 keV, 100-170 keV, 170-300 keV and 300-3000 keV for the low-gain data of the HXM.  In the SGM, 
40-100 keV, 100-230 keV, 230-450 keV and 450-1000 keV for the high-gain data and 560-840 keV, 840-1500 keV, 
1.5-2.6 MeV and 2.6-28 MeV for the low-gain data are investigated.  The highest SNR between $\pm$10 s window is 
4.7 at 7.5 s after the LIGO trigger in the 7-10 keV band of the HXM1 (the 1 s foreground and the 16 s background interval).  
Using 38,900 trials the false-detection probability at the level of 4.7 $\sigma$ was 
evaluated as $\sim$0.02 which is too high to claim a detection.  
In the HXM2 data, the SNR of the same time bin in which the highest SNR is found in HXM1 data is $-1.76$.  
The highest SNR is still found in the same time bin even if we extend the search window up to $\pm$60 s.  
Therefore, we concluded that there are no significant signals in the CGBM data associated with the 
gravitational-wave event.  Note that, however, our 
search is limited by the available continuous light curve data in the 0.125 s time resolution and might not be sensitive 
to an event with duration shorter than 0.125 s.  

\begin{figure}
\centerline{
\includegraphics[scale=0.45,angle=0]{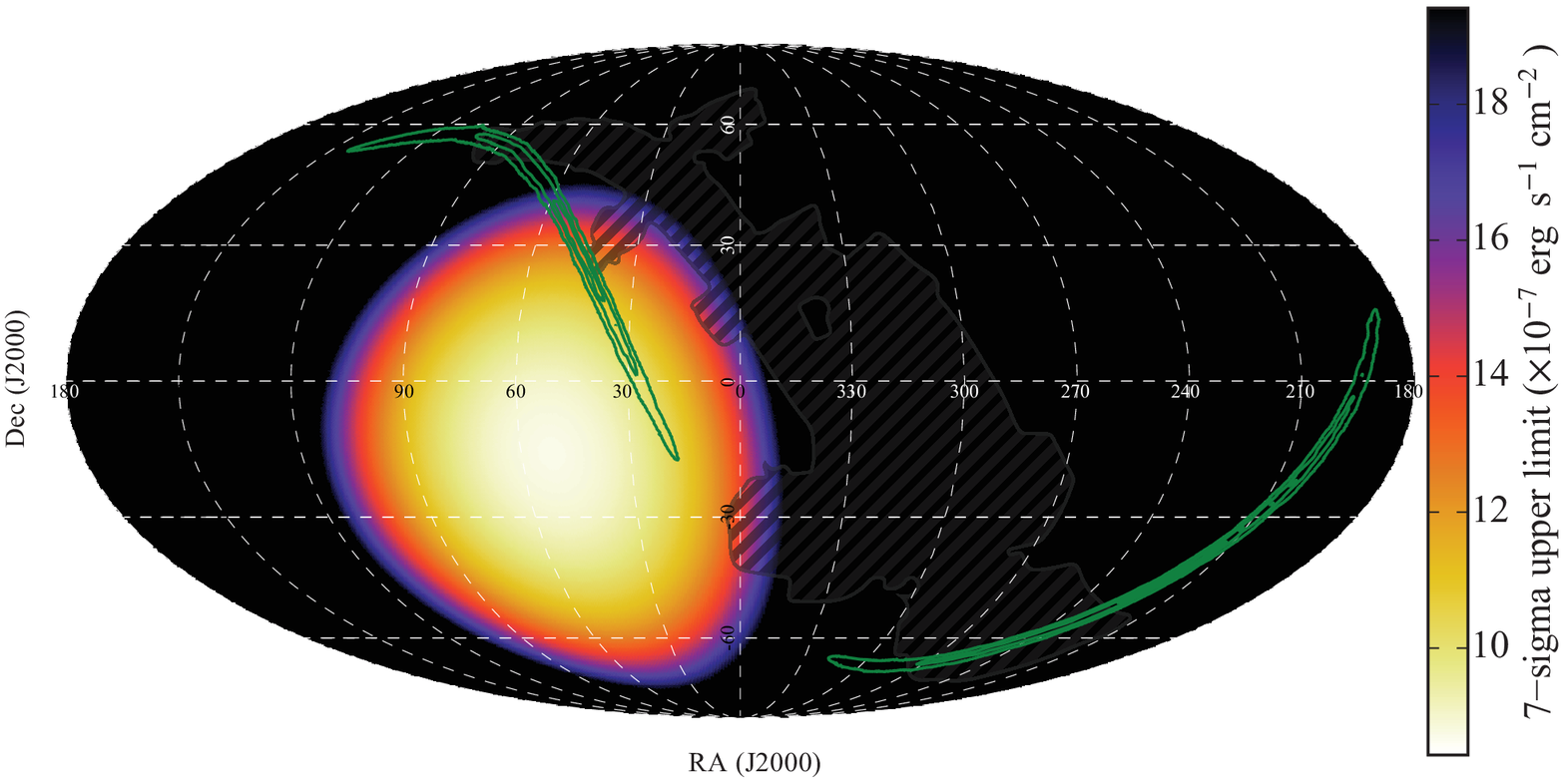}
\includegraphics[scale=0.45,angle=0]{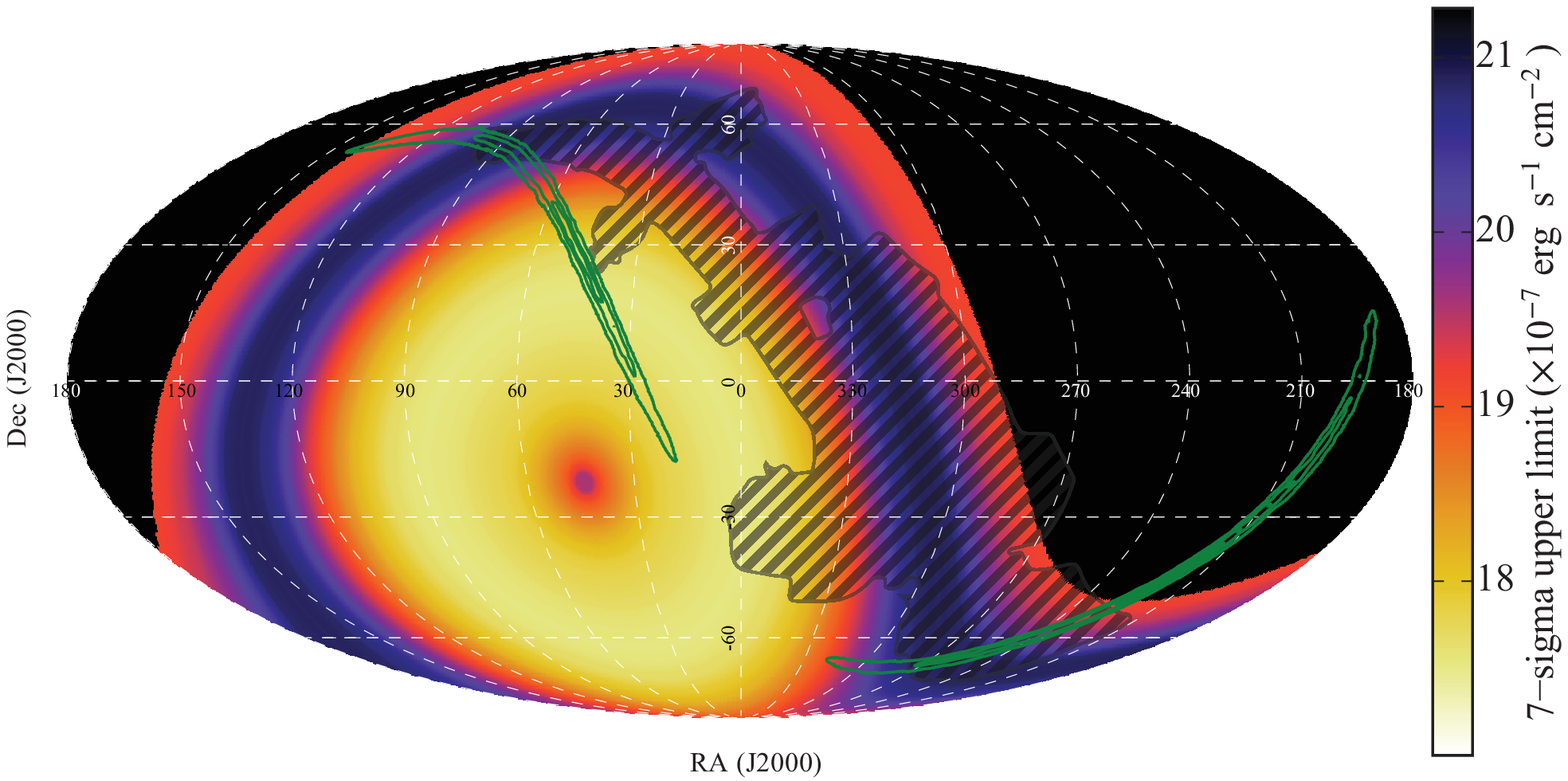}}
\caption{The sky maps of the 7 $\sigma$ upper limit for HXM (left) and SGM (right).  The assumed spectrum 
for estimating the upper limit is a typical BATSE S-GRBs (see text for details).  The energy bands are 7-500 keV 
for HXM and 50-1000 keV for SGM.    The GW 151226 probability map is shown in green contours.  The shadow of ISS is shown 
in black hatches.}
\label{fig:cgbm_sen}
\end{figure}

The flux upper limits of HXM and SGM are evaluated by using a CGBM Monte-Carlo simulator based on 
the {\it GEANT4} software package.  The simulations are performed by emitting the photons at incident 
angles from 0$^{\circ}$ to 110$^{\circ}$ in 1$^{\circ}$ steps with respect to the detector.  
The source spectrum assumes 
following two cases.  The first case is a typical GRB spectrum for the BATSE short GRBs (s-GRBs).  In this case, we use 
the averaged BATSE s-GRB spectral parameters in a cutoff power-law model\footnote{f(E) $\sim$ $E^{\alpha} \exp(-E\,(2+\alpha)/E_{\rm peak}$)}
reported by \citet{ghirlanda_batse_shortgrb}, with a photon index $\alpha$ of $-0.58$ and $E_{\rm peak}$ = 355 keV.  
The second case is the Crab-like spectrum: a power-law with a photon index of $-2.1$.  
The background spectrum is estimated using the real data over three days around the event in count space,  
normalized to the actual background level at the trigger time.  
The background variation was rather stable since the CGBM was not operated at the high 
background regions such as a high longitude and the South Atlantic Anomaly.  The gain differences during 
those three days were less than 3\% for both the HXM and the SGM data.  
The exposure time of the input and the background spectrum is one second.  
The source flux is evaluated to be in a range from $10^{-8}$ to $10^{-6}$ erg cm$^{-2}$ s$^{-1}$.   
The energy ranges for calculating the upper limits are determined as the best energy 
band to detect typical BATSE s-GRBs: 7-500 keV for HXM and 50-1000 keV for SGM.  
We also include the systematic uncertainties in the detector energy response function 
in the estimations of the upper limits of each detector.  
This systematic uncertainty is a correction factor of $\sim$2 in the effective 
area for taking into account the current calibration uncertainty at the incident angle between 
the on-axis and the far off-axis case.  
The sky maps of the 7 $\sigma$ upper limit overlaid with the shadow of ISS 
are shown in figure \ref{fig:cgbm_sen}.  
The 7 $\sigma$ threshold is the same setup parameter as the on-board trigger system.  
The upper limits assuming the typical BATSE s-GRB spectrum for the HXM and the SGM are 
1.0 $\times 10^{-6}$ erg cm$^{-2}$ s$^{-1}$ (7-500 keV) at the incident angle of 30$^{\circ}$ 
and 1.8 $\times 10^{-6}$ erg cm$^{-2}$ s$^{-1}$ (50-1000 keV) at 
the incident angle of 45$^{\circ}$, respectively.  
The incident angle of $\sim$$30^{\circ}$ of HXM corresponds to a half angle of the FOV from 
the boresight.  Whereas, SGM reaches to its maximum effective area at the incident angle of $\sim$$45^{\circ}$.  
In the case of the Crab-like spectrum, the 7 $\sigma$ upper limits of the HXM and the SGM are 
5.1 $\times 10^{-7}$ erg cm$^{-2}$ s$^{-1}$ (7-500 keV) 
at 30$^{\circ}$ off-axis and 1.4 $\times 10^{-6}$ erg cm$^{-2}$ s$^{-1}$ (50-1000 keV) at 45$^{\circ}$ off-axis. 

Our upper limits correspond to the k-corrected luminosity of $3.9 \times 10^{49}$ erg s$^{-1}$ for 
HXM and $4.7 \times 10^{49}$ erg s$^{-1}$ for SGM in the 1 keV - 10 MeV band at the rest frame using 
the luminosity distance of 440 Mpc and assuming a typical BATSE s-GRB spectrum.  
The isotropic-equivalent luminosity of s-GRBs is in the range from $5 \times 10^{48}$ erg s$^{-1}$ 
to $1 \times 10^{52}$ erg s$^{-1}$ with the mean of $1.6 \times 10^{51}$ erg s$^{-1}$ \citep{berger_2014}.  
Therefore, if s-GRBs occur within 440 Mpc, CGBM could 
detect a signal from a majority of s-GRBs. 

The CGBM coverage of the LIGO sky probability is estimated as follows.  First, we define the sky region 
by adding the probability of each pixel of the LIGO probability map (LALInference\_skymap\_2.fits) 
from the highest pixel until the summed 
probability reaches a 90\% level.  Then, the pixel values inside the overlapping region between 
this 90\% LIGO probability map and the FOV of CGBM are integrated to estimate the LIGO summed probabilities.  
Furthermore, the shadow due to the ISS structure is taken 
into account for the estimation of SGM.  The coverages of the summed LIGO probability are estimated as 
32.5\% for HXM and 49.1\% for SGM.  

\begin{table}[t]
\centering
\caption{Summary of the 7 $\sigma$ upper limits of the HXM and the SGM assuming the typical BATSE s-GRB 
and the Crab-like spectrum.} \label{tab:cgbm_ul}
\begin{tabular}{l|c|c}\hline
               & HXM (7-500 keV; 30$^{\circ}$ off-axis) & SGM (50-1000 keV; 45$^{\circ}$ off-axis)\\\hline\hline
s-GRB    &  $1.0 \times 10^{-6}$ erg cm$^{-2}$ s$^{-1}$ & $1.8 \times 10^{-6}$ erg cm$^{-2}$ s$^{-1}$\\
Crab-like &  $5.1 \times 10^{-7}$ erg cm$^{-2}$ s$^{-1}$ & $1.4 \times 10^{-6}$ erg cm$^{-2}$ s$^{-1}$\\\hline
\end{tabular}
\end{table}

\subsection{CAL Data Analysis and Results}\label{subsec:cal}

A search for gamma-ray events associated with GW 151226 was carried out using the CAL data 
in the time interval from $-525$ s to $+211$ s around the LIGO trigger.  The CAL was operational in low-energy 
gamma-ray mode 
in which the energy threshold is 1 GeV (compared to 10 GeV in high-energy mode) 
in this time period.  
We apply a gamma-ray selection by tracking pair creation events in the imaging calorimeter \citep{mori_2013}.  
The gamma-ray event selection used in this analysis is basically the same as the one of  \citet{mori_2013} 
although a stronger cut was applied by requiring three or more hits for track reconstruction.  
This ensures a higher tracking quality in exchange for a reduction of 1 radiation-length
in conversion materials (Tungsten) usable for pair creation in the imaging calorimeter.  
According to the simulation study which is generated events around the instrument isotropically, 
we estimate the highest gamma-ray efficiency is achieved around 10 GeV with an 
efficiency of 50\% relative to a geometrical factor of 420 cm$^2$ sr, which is the 100\% efficiency case, 
by applying the event selections described above.  
The effective areas for incident angles of 0$^{\circ}$, 20$^{\circ}$ and 30$^{\circ}$ are 
74 cm$^2$, 44 cm$^2$, and 17 cm$^2$ at 1 GeV, respectively.
The effective areas are increasing with energy and reach around 10 GeV 
their maxima of 260 cm$^2$, 180 cm$^2$ and 80 cm$^2$ 
for incident angles of 0$^{\circ}$, 20$^{\circ}$ and 30$^{\circ}$, respectively.  
Our long-term CAL observation of galactic diffuse gamma-rays in the low-energy gamma-ray mode
clearly identified a peak at the galactic equator on the count map as a function of the galactic latitude.  
This matches the expectation estimated based on a galactic diffuse 
radiation model \citep{lat_diffusegamma} 
when considering above mentioned effective areas and observation exposure.  
As a result, it was proven that the {\it CALET} observation in low-energy gamma-ray 
mode has achieved detection of the galactic diffuse gamma-rays.
Since the searched location for the GW 151226 counterpart is significantly 
far from the galactic plane, the number of background gamma-rays is negligibly small, 0.0024 
events according to the calculation based on the model of \citet{lat_diffusegamma}.  
Another expected background might, however, result from misidentification of
cosmic-ray events at lower energies. The number of such events in the
time-window of the GW 151226 counterpart search is also estimated using
the diffuse gamma-ray model in comparison with the observed data.
A conservative upper limit of this background is obtained by the assumption
that all of the excess in observed data to the model originates from 
the background.  Then, the possibility of such a
misidentification is confirmed to be less than 0.035 events.  Therefore,
the CAL observation is virtually background free in such a short time window.  
We found no gamma-ray candidate inside this time window with negligible 
contamination from other events.  

\begin{figure}
\centerline{
\includegraphics[scale=0.45,angle=0]{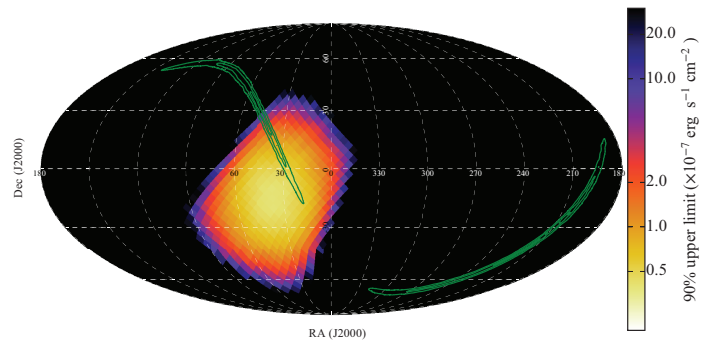}}
\caption{The sky map of the 90\% upper limit for CAL in the 1-100 GeV band.  A power-law model 
with a photon index of $-2$ is used to calculate the upper limit.  The GW 151226 probability map is shown in green contours.}
\label{fig:cal_sen}
\end{figure}

The upper limit of the CAL observation in this 736 s long period is estimated as follows.
First, we calculated the effective area and the resultant exposure map in the time window for the 1-100 GeV band.  
At lower energies, the effective area gradually decreases below 10 GeV and reaches zero around 500 MeV.  
Next, we estimated the limiting flux corresponding to 2.44 events, which is the 90\% confidence limit for null observation, 
assuming a single power-law model with a photon index of $-2$ by applying the estimated exposure map.  
The assumed photon index of $-2$ is a typical photon index of the {\it Fermi}-LAT GRBs in the GeV energy range \citep{lat_grb_cat}.  
Figure \ref{fig:cal_sen} shows the sky map of the flux upper limit at 90\% confidence level.  The estimated 90\% 
upper limit is $2 \times 10^{-7}$ erg cm$^{-2}$ s$^{-1}$ in the 1 - 100 GeV band where CAL reaches 
15\% of the integrated LIGO probability ($\sim$1.1 sr).  The CAL upper limit in luminosity 
is estimated as $4.6 \times 10^{48}$ erg s$^{-1}$ for a source distance of 440 Mpc. 
The flux upper limit in the same energy band as reported by {\it Fermi}-LAT of 0.1-1 GeV \citep{racusin2016} 
is calculated to be $1 \times 10^{-7}$ erg cm$^{-2}$ s$^{-1}$ extrapolating a single power-law spectrum 
with a photon index of $-2$.  

\vspace{0.3cm}
We would like to thank the anonymous referee for comments and suggestions that materially improved the paper.  We 
gratefully acknowledge JAXA's contributions for {\it CALET} development and operation on ISS.  We express our sincere thanks to 
ASI and NASA for their support to the {\it CALET} project.  This work is partially supported by JSPS Grant-in-Aid for Scientific Research (S) 
Number 26220708 and MEXT-Supported Program for the Strategic Research Foundation at Private Universities (2011-2015) S1101021 
in Waseda University.  This work is also supported in part by MEXT Grant-in-Aid for Scientific Research on Innovative Areas  
Number 24103002.  
US CALET work is supported by NASA under RTOP 14-APRA14-0075 (GSFC) and grants NNX16AC02G (WUSL), NNX16AB99G (LSU), 
and NNX11AE06G (Denver).

\newpage

\newpage

\newpage

\end{document}